# Towards Post-Quantum Blockchain: A Review on Blockchain Cryptography Resistant to Quantum Computing Attacks

**TIAGO M. FERNÁNDEZ-CARAMÉS[1], (Senior Member, IEEE),**
**AND PAULA FRAGA-LAMAS[1], (Member, IEEE)**
[1] Department of Computer Engineering, Faculty of Computer Science, Centro de Investigación CITIC, Campus de Elviña s/n, Universidade da Coruña, 15071, A Coruña, Spain.
(e-mail: tiago.fernandez@udc.es; paula.fraga@udc.es).

Corresponding authors: Tiago M. Fernández-Caramés and Paula Fraga-Lamas (e-mail: tiago.fernandez@udc.es; paula.fraga@udc.es).

This work has been funded by the Agencia Estatal de Investigación of Spain (TEC2016-75067-C4-1-R) and ERDF funds of the EU (AEI/FEDER, UE).

**ABSTRACT** Blockchain and other Distributed Ledger Technologies (DLTs) have evolved significantly in the last years and their use has been suggested for numerous applications due to their ability to provide transparency, redundancy and accountability. In the case of blockchain, such characteristics are provided through public-key cryptography and hash functions. However, the fast progress of quantum computing has opened the possibility of performing attacks based on Grover's and Shor's algorithms in the near future. Such algorithms threaten both public-key cryptography and hash functions, forcing to redesign blockchains to make use of cryptosystems that withstand quantum attacks, thus creating which are known as post-quantum, quantum-proof, quantum-safe or quantum-resistant cryptosystems. For such a purpose, this article first studies current state of the art on post-quantum cryptosystems and how they can be applied to blockchains and DLTs. Moreover, the most relevant post-quantum blockchain systems are studied, as well as their main challenges. Furthermore, extensive comparisons are provided on the characteristics and performance of the most promising post-quantum public-key encryption and digital signature schemes for blockchains. Thus, this article seeks to provide a broad view and useful guidelines on post-quantum blockchain security to future blockchain researchers and developers.

## I. INTRODUCTION

Blockchain is a technology that was born with the cryptocurrency Bitcoin [1] and that is able to provide secure communications, data privacy, resilience and transparency [2]. A blockchain acts as a distributed ledger based on a chain of data blocks linked by hashes that allow for sharing information among peers that do not necessarily trust each other, thus providing a solution for the double-spending problem [3], [4], [5]. Such features have popularized blockchain in the last years and it has already been suggested as a key technology for different applications related to smart health [6], measuring systems [7], logistics [8], [9], e-voting [10] or smart factories [11], [12].

Blockchain users interact securely with the blockchain by leveraging public-key/asymmetric cryptography, which is essential for authenticating transactions. Hash functions are also key in a blockchain, since they allow for generating digital signatures and for linking the blocks of a blockchain. The problem is that both public-key cryptosystems and hash functions are threatened by the evolution of quantum computers. In the case of public-key cryptosystems, secure transaction data may be recovered fast by future quantum computing attacks. Such attacks impact the most popular public-key algorithms, including RSA (Rivest, Shamir, Adleman) [13],

ECDSA (Elliptic Curve Digital Signature Algorithm) [14], [15], ECDH (Elliptic Curve Diffie-Hellman) [16] or DSA (Digital Signature Algorithm) [17], which can be broken in polynomial-time with Shor's algorithm [18] on a sufficiently powerful quantum computer. Moreover, quantum computers can make use of Grover's algorithm [19] to accelerate the generation of hashes, which enables recreating the entire blockchain. Furthermore, Grover's algorithm may be adapted to detect hash collisions, which can be used to replace blocks of a blockchain while preserving its integrity.

This article analyzes how to evolve blockchain cryptography (i.e., its public-key security algorithms and hash functions) so that it can resist quantum computing attacks based on Grover's and Shor's algorithms, thus deriving into the creation of post-quantum blockchains. To guide researchers on the development of such a kind of blockchains, this article first provides a broad view on the current state of the art of post-quantum cryptosystems. Specifically, the most relevant post-quantum cryptosystems for blockchains are analyzed, as well as their main challenges. Furthermore, extensive comparisons are provided on the characteristics and performance of the most promising post-quantum public-key encryption and digital signature schemes.

The rest of this article is structured as follows. Section II describes the essential concepts related to blockchain and to its security primitives. Section III studies the impact of quantum attacks on blockchain public-key security schemes and on the most popular hash functions. In addition, Section III enumerates the most relevant post-quantum initiatives, emphasizing the ones related to blockchain and indicating the main features that a blockchain post-quantum scheme would need to provide. Section IV reviews the main types of post-quantum public-key and digital signature schemes, and analyzes their application to blockchain. Section V studies the performance of the most promising post-quantum cryptosystems when running them on hardware that can be used by blockchain nodes. Section VI details the main blockchain proposals that have already considered the use of post-quantum schemes. Section VII indicates the most significant challenges currently posed by post-quantum blockchain schemes and points at different paths to be followed by future researchers and developers. Finally, Section VIII summarizes the most relevant findings of this review article and Section IX is dedicated to conclusions.

## II. BLOCKCHAIN BASICS AND CRYPTOGRAPHIC PRIMITIVES
### A. TERMINOLOGY AND KEY CONCEPTS

Before starting to review the state of the art on post-quantum blockchains (i.e., on blockchains whose cryptosystems can resist quantum computing attacks), it is necessary to introduce several basic concepts, since some of the terminology may vary in the literature from one author to another.

It is first important to note that the concept of blockchain has evolved significantly since its original definition for Bitcoin [1]. In fact, researchers are still discussing the different elements that a blockchain has to contemplate to be actually considered a blockchain. The most common definition of blockchain is the one given in the Introduction of this article: it is a public ledger that stores data (e.g., transaction information, an event log) that are shared among multiple entities that do not necessarily trust each other. Every transaction on the blockchain is verified and stored by following a consensus protocol. Once a transaction is stored, ideally, it cannot be removed from the blockchain without making a significant computational effort.

A blockchain node is a computational entity able to perform operations on the blockchain. It is common to distinguish between regular blockchain nodes, which only interact with the blockchain, and full nodes, which have a copy of the blockchain and contribute to it by validating transactions. A blockchain miner is a third type of node that is present in many blockchains and whose contribution is essential during blockchain transaction validations: to carry out the validation, they perform certain actions following a consensus protocol. There are many consensus protocols [20], being some of the most popular Proof-of-Work (PoW) (used by Bitcoin), the variants of the Byzantine Fault Tolerance (BFT) methods [21] or Proof-of-Stake (PoS).

The concept of smart contract is also relevant: it is a piece of code stored on the blockchain that can be executed autonomously. Smart contracts can be used to automate certain tasks depending on the state of the blockchain and in other external data sources called oracles [22].

The previously introduced concepts have contributed to the success of blockchain and to its main security features:

- *i* Decentralization. If one node of the blockchain is attacked or shut down, its information keeps on being available from the other blockchain nodes.
- *i* Data privacy and integrity. Blockchain uses public-key cryptography and hash functions for providing data privacy, integrity and authentication.
- *i* Data immutability. Once a transaction is stored on the blockchain, it is not possible to make further modifications on it (the only exception is blockchain forks [2], which require to reach a consensus among the entities that participate on the blockchain).

A detailed description on the inner workings of the previously mentioned blockchain components and algorithms is out of the scope of this paper, but the interested reader can find further information in [2], [22], [23], [24], [25], [26], [27], [28].

### B. BLOCKCHAIN SECURITY PRIMITIVES
The security features provided by blockchain are essentially sustained by public-key/asymmetric cryptography and hash functions, whose role in blockchain security is detailed in the next subsections.

#### 1) Public-Key cryptography
A blockchain usually makes use of public-key cryptosystems for securing information exchanges between parties by au-thenticating transactions through digital signatures. During the signature process, the signer signs with a private key, while the public key, which is shared publicly, is used to verify that the signature is valid. Thus, when a signing algorithm is secure, it is guaranteed that only the person with a private key could have generated certain signature. For instance, Bitcoin uses ECDSA signatures with the Koblitz curve secp256k1, which depends on a private key for signing messages and on the corresponding public key for checking the signature.

Public-key cryptography is also essential for the so-called wallets, which are private key containers that store files and simple data. Thus, in a blockchain system each user has a wallet that is associated with at least a public address (usually a hash of the user public key) and a private key that the user needs for signing transactions. For instance, in blockchains like Bitcoin every transaction ends up being 'sent' to the public address of the receiver and is signed with the private key of the sender. In order to spend bitcoins, their owner has to demonstrate the ownership of a private key. To verify the authenticity of the received currency, every entity that receives bitcoins verifies its digital signature by using the public key of the sender.

#### 2) Hash functions

Hash functions like SHA-256 or Scrypt are commonly used by blockchains because they are easy to check, but really difficult to forge, thus allowing the generation of digital signatures that blockchain users need to authenticate themselves or their data transactions in front of others.

Hash functions are also used by blockchains to link their blocks (i.e., groups of transactions that are considered to occur at the same time instant). Such blocks are linked in chronological order, containing each block the hash of the previous block. It is straightforward to hash a block of a blockchain, but some blockchains like Bitcoin restrict block hashing to make it meet a specific mathematical condition (e.g., the hash should contain a number of leading zeros [1]), which slows down block addition.

Finally, it is worth mentioning that hash functions are used in blockchains for generating user addresses (i.e., user public/private keys) or for shortening the size of public addresses [29], [30].

### III. FROM PRE-QUANTUM TO POST-QUANTUM BLOCKCHAIN
#### A. BLOCKCHAIN PUBLIC-KEY SECURITY

It must be first noted that public-key cryptosystems strength against classical computing attacks has been traditionally estimated through the so-called bits-of-security level. Such a level is defined as the effort required by a classical computer to perform a brute-force attack. For instance, an asymmetric cryptosystem has a 1024-bit security when the effort required to attack it with a classical computer is similar to the one needed to carry out a brute-force attack on a 1024-bit cryptographic key. As a reference, Table 1 indicates the security level of some of the most popular symmetric and asymmetric cryptosystems.

**TABLE 1.** Reference security levels for popular symmetric and asymmetric cryptosystems (source: [31]).

| Security Level | Symmetric Cryptosystem Key Size | RSA Key Size | ECDSA Curve Key Size |
|---|---|---|---|
| 80 | 2TDEA (112 bits) | 1024 bits | prime192v1 (192 bits) |
| 112 | 3TDEA (168 bits) | 2048 bits | secp224r1 (224 bits) |
| 128 | AES-128 (128 bits) | 3072 bits | secp256r1 (256 bits) |
| 192 | AES-192 (192 bits) | 7680 bits | secp384r1 (384 bits) |

The cost of breaking current 80-bit security cryptosystems with classical computers is estimated to be between tens of thousands and hundreds of millions of dollars. In the case of 112-bit cryptosystems, they are considered to be secure to classical computing attacks for the next 30 to 40 years [32]. However, researchers have determined that 160-bit elliptic curves can be broken with a 1000-qubit quantum computer, while 1024-bit RSA would need roughly 2,000 qubits [33]. Such a threat affects not only cryptosystems that rely on integer factorization (e.g., RSA) or elliptic curves (e.g., ECDSA, ECDH), but also others based on problems like the discrete logarithm problem [34], which can be solved fast through Shor's algorithm.

As of writing, powerful quantum computers are not available: the most powerful quantum computer (claimed by IonQ) has only 79 qubits and even technologically-advanced organizations like the U.S. National Security Agency (NSA) seem to have not made significant progress on large quantum computers [35]. However, it is estimated that in the next 20 years such a kind of computers will be functional enough to be able to break easily current strong public-key cryptosystems [36]. In fact, organizations like the NSA have already warned on the impact of quantum computing on IT products and recommended increasing the ECC (Elliptic Curve Cryptography) security level of certain cryptographic suites [34]. Although some researchers have speculated on the real reasons behind such an NSA announcement [37], long-term public-key cryptography seems to be threatened and developers need to prepare current blockchains for the post-quantum era.

Table 2 indicates the main characteristics of the most relevant public-key cryptosystems that are affected by the quantum threat. The Table also includes the characteristics of other relevant cryptosystems that will be broken or that will be impacted significantly by quantum attacks related to Shor's and Grover's algorithms.

#### B. HASH FUNCTION SECURITY

In contrast to public-key cryptosystems, traditional hash functions are considered to be able to withstand quantum attacks since it seems unlikely the development of quantum algorithms for NP-hard problems [38]. Although new hash functions have been recently proposed by academics to resist quantum attacks [39], it is usually recommended to increase the output size of traditional hash functions. This recommendation is related to quantum attacks that can make use of Grover's algorithm to accelerate brute force attacks by a quadratic factor [36]. Specifically, Grover's algorithm can be used in two ways to attack a blockchain:

- First, to search for hash collisions and then replace entire blockchain blocks. For instance, in the specific case of the work described in [41], it is proposed to use Grover's algorithm to find collisions in hash functions, concluding that a hash function would have to output $3*n$ bits to provide a n-bit security level. Such a conclusion means that many current hash functions would not be valid for the post-quantum era, while others like SHA-2 or SHA-3 will have to increase their output size.
- Second, Grover's algorithm can be used to accelerate mining in blockchains like Bitcoin (i.e., it is able to speed up the generation of nonces), which would allow for recreating entire blockchains fast, thus undermining their integrity.

**TABLE 2.** Main blockchain and popular cryptosystems impacted by the quantum threat.

| Algorithm | Main Affected Blockchains/DLTs | Function | Pre-Quantum Security Level | Estimated Post-Quantum Security Level | Key Size | Hash/Signature Size |
|---|---|---|---|---|---|---|
| SHA-256 | Bitcoin, Ethereum, Dash, Litecoin, Zcash, Monero, Ripple, NXT, Byteball | Hash function | 256 bits | 128 bits (Grover) | - | 256 bits |
| Ethash (Keccak-256, Keccak-512) | Ethereum | Hash function | 256/512 bits | 128/256 bits (Grover) | - | 256/512 bits |
| Scrypt | Litecoin, NXT | Hash function | 256 bits | 128 bits (Grover) | - | 256 bits |
| RIPEMD160 | Bitcoin, Ethereum, Litecoin, Monero, Ripple, Bytecoin | Hash function | 160 bits | 80 bits (Grover) | - | 160 bits |
| Keccak-256 | Monero, Bytecoin | Hash function | 256 bits | 128 bits (Grover) | - | 256 bits |
| Keccak-384 | IOTA | Hash function | 384 bits | 192 bits (Grover) | - | 384 bits |
| ECDSA | Bitcoin, Ethereum, Dash, Litecoin, Zcash, Ripple, Byteball | Signature | 128 bits | Broken (Shor) | 256 bits | 520 bits |
| RSA-1024 | - | Signature, Encryption | 80 bits | Broken (Shor) | 1024 bits | 1024 bits |
| RSA-2048 | - | Signature, Encryption | 112 bits | Broken (Shor) | 2048 bits | 2048 bits |
| RSA-3072 | - | Signature, Encryption | 128 bits | Broken (Shor) | 3072 bits | 3072 bits |
| DSA-3072 | - | Signature | 128 bits | Broken (Shor) | 3072 bits | - |
| SHA-3 256 | - | Hash function | 256 bits | 128 bits (Grover) | - | 256 bits |
| AES-128 | - | Symmetric Encryption | 128 bits | 64 bits (Grover) | 128 bits | - |
| AES-256 | - | Symmetric Encryption | 256 bits | 128 bits (Grover) | 256 bits | - |

In addition, quantum attacks through Shor's algorithm also impact hash functions: if a blockchain hash function is broken, someone with a powerful enough quantum computer may use Shor's algorithm to forge digital signatures, to impersonate blockchain users and to steal their digital assets. As a reference, Table 2 includes the main characteristics of the most popular hash functions that are currently used by relevant blockchains and indicates the impact of quantum computing on their security level.

### C. POST-QUANTUM BLOCKCHAIN INITIATIVES

Post-quantum cryptography is currently a hot topic that has been addressed by research projects (e.g., PQCrypto [42], SAFEcrypto [43], CryptoMathCREST [44] or PROMETHEUS [45]), standardization initiatives [46], [47], [48], [49], [50], [51], [52], [53] and workshops [54], [55], [56], which obtained relevant results [57], [58], [59] and produced interesting reports [32], [60], [61], [62], [63], [64], [65]. Among the previously mentioned initiatives, it is worth noting the NIST call for proposals for post-quantum public-key cryptosystems [66], which is currently in its second round [67] and which is expected to deliver the first standard drafts between 2022 and 2024.

Although the previous projects and initiatives generated very valuable results, they were not explicitly focused on post-quantum blockchains. However, there have been specific post-quantum initiatives related to the most popular blockchains. For instance, Bitcoin Post-Quantum is an experimental branch of Bitcoin's main blockchain that uses a post-quantum digital signature scheme [68]. Another example is Ethereum 3.0, which plans to include quantum-resistant components like zk-STARKs (Zero-Knowledge Scalable Transparent ARguments of Knowledge) [69]. Other blockchain platforms like Abelian [70] have suggested using lattice-based post-quantum cryptosystems to prevent quantum attacks, while certain blockchains such as Corda are experimenting with post-quantum algorithms like SPHINCS [71].

### D. IDEAL CHARACTERISTICS OF BLOCKCHAIN POST-QUANTUM SCHEMES

In order to be efficient, a post-quantum cryptosystem would need to provide blockchains with the following main features:

- Small key sizes. The devices that interact with a blockchain need to ideally make use of small public and private keys in order to reduce the required storage space. In addition, small keys involve less complex computational operations when managing them. This is especially important for blockchains that require the interaction of Internet of Things (IoT) end-devices, which are usually constrained in terms of storage and computational power. It is worth indicating that IoT, like other emerging technologies (e.g., deep learning [72]), has experienced a significant growth in the last years [73], [74], [75], [76], [77], but IoT devices still face some important challenges, mainly regarding security [78], [79], [80], [81], [82], which are limiting to some extent its jointly use with blockchain and its widespread adoption.
- Small signature and hash length. A blockchain essentially stores data transactions, including user signatures and data/block hashes. Therefore, if signature/hash length increases, blockchain size will also increase as well.
- Fast execution. Post-quantum schemes need to be as fast as possible in order to allow a blockchain to process a large amount of transactions per second. Moreover, a fast execution usually involves low computational complexity, which is necessary to not to exclude resource-constrained devices from blockchain transactions.
- Low computational complexity. This feature is related to a fast execution, but it is important to note that a fast execution with certain hardware does not imply that the post-quantum cryptosystem is computationally simple. For instance, some schemes can be executed fast in Intel microprocessors that make use of the Advanced Vector Extensions 2 (AVX2) instruction set, but the same schemes may be qualified as slow when executed on ARM-based microcontrollers. Therefore, it is necessary to look for a trade-off between computational complexity, execution time and supported hardware devices.
- Low energy consumption. Some blockchains like Bitcoin are considered to be power hungry mainly due to the energy required to execute its consensus protocol. There are other factors that impact power consumption, like the used hardware, the amount of performed communications transactions and, obviously, the implemented security schemes, which can draw a relevant amount of current due to the complexity of the performed operations [83], [84].

## IV. POST-QUANTUM CRYPTOSYSTEMS FOR BLOCKCHAIN

There are four main types of post-quantum cryptosystems and a fifth kind that actually mixes both pre-quantum and post-quantum cryptosystems. The following subsections analyze the potential application of such schemes for the implementation of encryption/decryption mechanisms and for signing blockchain transactions.

A detailed description on the algorithms cited in the next subsections is out of the scope of this article, but the interested reader can consult the specific references cited throughout the text and books like [85], which provide a wide but comprehensive description of the most popular post-quantum cryptosystems.

As a summary, the five different types of post-quantum cryptosystems are depicted in Figure 1 together with examples of encryption and digital signature scheme implementations.

### A. PUBLIC-KEY POST-QUANTUM CRYPTOSYSTEMS

#### 1) Code-based cryptosystems

They are essentially based on the theory that supports error-correction codes. For instance, McEliece's cryptosystem is an example of code-based cryptosystem [86] that dates back from the 70s and whose security is based on the syndrome decoding problem [87]. McEliece's scheme provides fast encryption and relatively fast decryption, which is an advantage for performing rapid blockchain transactions. However, McEliece's cryptosystem requires to store and perform operations with large matrices that act as public and private keys. Such matrices usually occupy between 100 kilobytes and several megabytes, which may be a restriction when resource-constrained devices are involved. To tackle this issue, future researchers will have to study matrix compression techniques, as well as the use of different codes (e.g., Low-Density Parity-Check (LDPC) codes, Quasi-Cyclic Low-Rank Parity-Check (QC-LRPC) codes) and specific coding techniques [88].

As a reference, Table 3 compares the main characteristics of the public-key code-based post-quantum encryption cryptosystems that passed to the second round of the NIST call. There are other post-quantum cryptosystems [89], but the NIST second-round candidates are specially interesting due to their standardization chances and because they have been already thoroughly analyzed by the cryptographic community.

It is important to note that the parameters of the algorithms compared in Table 3 can be adjusted according to the required security and thus key size and performance may vary among them. Specifically, the cryptosystems of the Table were selected with the objective of comparing the ones with the smallest key sizes that provided the main quantum security levels demanded by NIST (128, 192 and 256 bits). The same criteria were applied for the selection of the algorithms compared in the rest of this article.

As it can be observed in Table 3, the evaluated code-based cryptographic schemes provide between 128 and 256 bits of classical security, but such a level is reduced significantly in terms of quantum security. Regarding the compared public/private key sizes, they range between very small sizes (320 bits, for the private keys of ROLLO-II and RQC) and up to 15.5 KB (for the public key of the highest security level of HQC). On average, even when making use of compression techniques, the size of code-based scheme keys is clearly larger than the one required by current ECDSA and RSA-based encryption systems.

It is worth pointing out that in the case of HQC two key sizes are indicated: the one inside parentheses is related to the use of a seed expander. However, note that during the execution of the algorithm an expanded key will consume the amount of memory indicated outside the parentheses and will also need to perform the expansion operation, which slows down the execution of the algorithm.

Overall, among the schemes compared in Table 3, it seems that RQC-II provides the best trade-off between security and key size, although it is not among the fastest post-quantum schemes (the performance of the algorithms in Table 3 is analyzed later in Section V).

#### 2) Multivariate-based cryptosystems

Multivariate-based schemes rely on the complexity of solving systems of multivariate equations, which have been demonstrated to be NP-hard or NP-complete [85]. Despite their resistance to quantum attacks, it is necessary further research for improving their decryption speed (due to the involved "guess work") and to reduce their large key size and ciphertext overhead [90].

Currently, some of the most promising multivariate-based schemes are the ones based on the use of square matrices with random quadratic polynomials, the cryptosystems derived from Matsumoto-Imai's algorithm and the schemes that rely on Hidden Field Equations (HFE) [91], [92], [93].

#### 3) Lattice-based cryptosystems

This kind of cryptographic schemes are based on lattices, which are sets of points in n-dimensional spaces with a periodic structure. Lattice-based security schemes rely on the presumed hardness of lattice problems like the Shortest Vector Problem (SVP), which is an NP-hard problem whose objective is to find the shortest non-zero vector within a lattice. There are other similar lattice-related problems like the Closest Vector Problem (CVP) or the Shortest Independent Vectors Problem (SIVP) [94], which nowadays cannot be solved efficiently through quantum computers.

Lattice-based schemes provide implementations that allow for speeding up blockchain user transactions since they are often computationally simple, so they can be executed fast and in an efficient way. However, like it occurs with other post-quantum schemes, lattice-based implementations need to store and make use of large keys, and involve large ciphertext overheads. For example, lattice-based schemes like

**TABLE 3.** Post-quantum code-based public-key encryption schemes that passed to the second round of the NIST call.

| Cryptosystem | Subtype | Claimed Quantum Security | Claimed Classical Security | Public Key Size (bits) | Private Key Size (bits) | Key References |
|---|---|---|---|---|---|---|
| BIKE-1 Level 1 | QC-MDPC McEliece | - | 128 bits | 20,326 | 2,130 | [95], [96], [97], [98], [99], [100], [101] |
| BIKE-1 Level 3 | QC-MDPC McEliece | - | 192 bits | 39,706 | 3,090 | [95], [96], [97], [98], [99], [100], [101] |
| BIKE-1 Level 5 | QC-MDPC McEliece | - | 256 bits | 65,498 | 4,384 | [95], [96], [97], [98], [99], [100], [101] |
| BIKE-2 Level 1 | QC-MDPC Niederreiter | - | 128 bits | 10,163 | 2,130 | [95], [96], [97], [98], [99], [100], [101] |
| BIKE-2 Level 3 | QC-MDPC Niederreiter | - | 192 bits | 19,853 | 3,090 | [95], [96], [97], [98], [99], [100], [101] |
| BIKE-2 Level 5 | QC-MDPC Niederreiter | - | 256 bits | 32,749 | 4,384 | [95], [96], [97], [98], [99], [100], [101] |
| BIKE-3 Level 1 | QC-MDPC Ouroboros | - | 128 bits | 22,054 | 2,010 | [95], [96], [97], [98], [99], [100], [101] |
| BIKE-3 Level 3 | QC-MDPC Niederreiter | - | 192 bits | 43,366 | 2,970 | [95], [96], [97], [98], [99], [100], [101] |
| BIKE-3 Level 5 | QC-MDPC Niederreiter | - | 256 bits | 72,262 | 4,256 | [95], [96], [97], [98], [99], [100], [101] |
| Classic McEliece (mceliece8192128) | Niederreiter's dual version using binary Goppa codes | - | 256 bits | 10,862,592 | 112,640 | [102], [103], [104], [105] |
| HQC Level 1 (hqc-128-1) | Quasi-Cyclic and BCH codes | 64 bits | 128 bits | 49,360 (25,000) | 2,016 (320) | [106], [107], [108] |
| HQC Level 3 (hqc-192-1) | Quasi-Cyclic and BCH codes | 96 bits | 192 bits | 87,344 (43,992) | 3,232 (320) | [106], [107], [108] |
| HQC Level 5 (hqc-256-1) | Quasi-Cyclic and BCH codes | 128 bits | 256 bits | 127,184 (63,912) | 4,256 (320) | [106], [107], [108] |
| LEDACrypt KEM Level 1 (for two circulant blocks) | QC-LDPC Niederreiter | - | 128 bits | 14,976 | 3,616 (192) | [109], [110], [111] |
| LEDACrypt KEM Level 3 (for two circulant blocks) | QC-LDPC Niederreiter | - | 192 bits | 25,728 | 5,152 (256) | [109], [110], [111] |
| LEDACrypt KEM Level 5 (for two circulant blocks) | QC-LDPC Niederreiter | - | 256 bits | 36,928 | 6,112 (320) | [109], [110], [111] |
| NTS-KEM Level 1 | Based on McEliece and Niederreiter | 64 bits | 128 bits | 2,555,904 | 73,984 | [112], [113] |
| NTS-KEM Level 3 | Based on McEliece and Niederreiter | 96 bits | 192 bits | 7,438,080 | 140,448 | [112], [113] |
| NTS-KEM Level 5 | Based on McEliece and Niederreiter | 128 bits | 256 bits | 11,357,632 | 159,376 | [112], [113] |
| ROLLO-II 128 | Based on rank metric codes with LRPC codes | - | 128 bits | 12,368 | 320 | [114], [115] |
| ROLLO-II 192 | Based on rank metric codes with LRPC codes | - | 192 bits | 16,160 | 320 | [114], [115] |
| ROLLO-II 256 | Based on rank metric codes with LRPC codes | - | 256 bits | 19,944 | 320 | [114], [115] |
| RQC-I | Based on Rank Quasi-Cyclic codes | - | 128 bits | 6,824 | 320 | [108], [116], [117] |
| RQC-II | Based on Rank Quasi-Cyclic codes | - | 192 bits | 11,128 | 320 | [108], [116], [117] |
| RQC-III | Based on Rank Quasi-Cyclic codes | - | 256 bits | 18,272 | 320 | [108], [116], [117] |

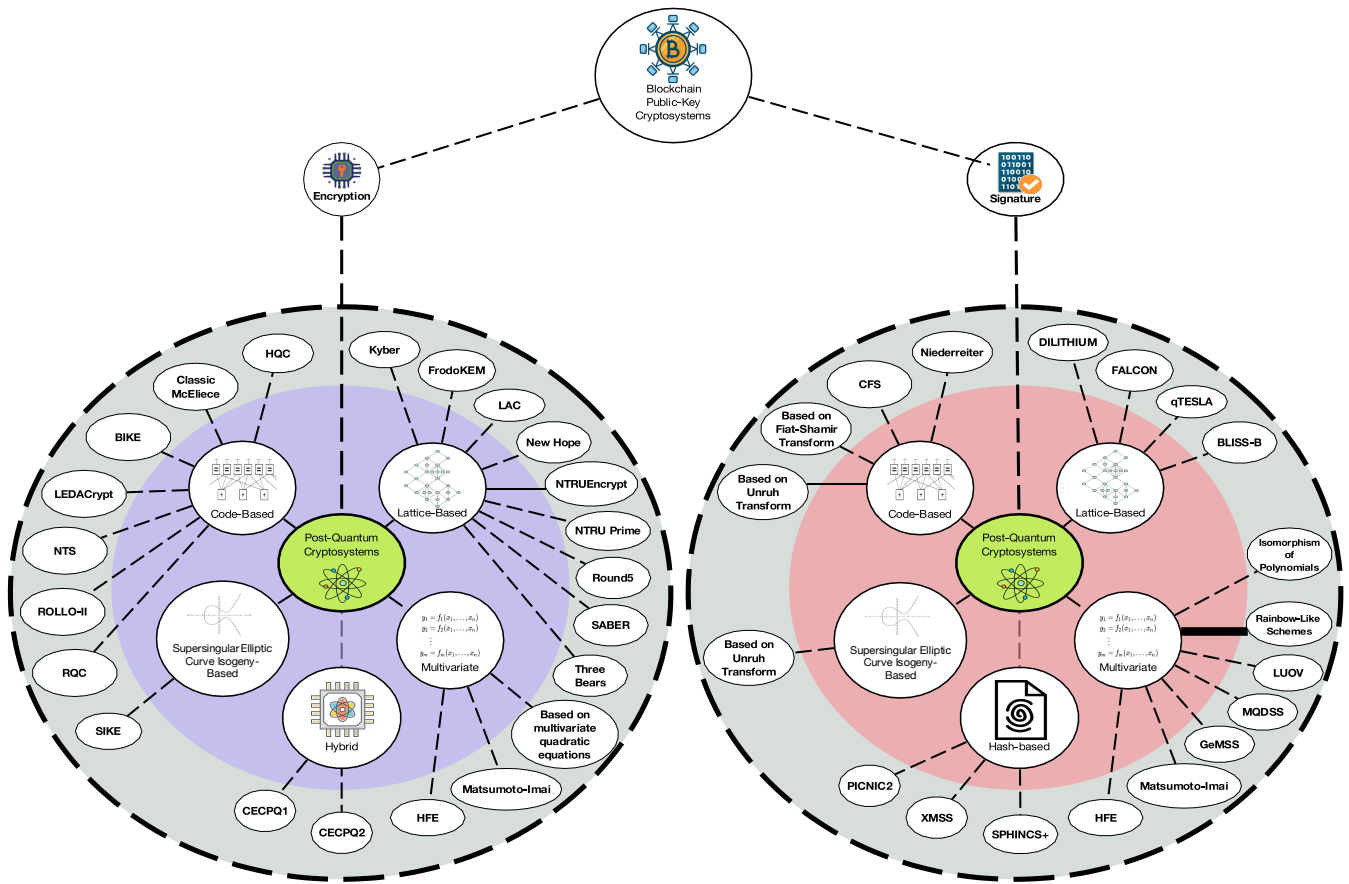

**FIGURE 1.** Post-quantum public-key cryptosystem taxonomy and main practical implementations.

NTRU [118] or NewHope [119] often require to manage keys in the order of a few thousand bits.

As of writing, the most promising lattice-based cryptosystems are based on polynomial algebra [118], [120], [121] and on the Learning With Errors (LWE) problem and its variants (e.g., LP-LWE (Lindner-Peikert LWE) or Ring-LWE [122], [123]).

Table 4 compares the public-key lattice cryptosystems that passed to the second round of the NIST call. As it can be observed in the Table, the included schemes provide a classical security between 128 and 368 bits and a quantum security between 84 and 300 bits, so their complexity differs significantly depending on the algorithm and on the provided security level. Key size also fluctuates remarkably: from the 128-bit private key of the IoT version of Round5, to the 344,704-bit private key of FrodoKEM-1344. As it was previously mentioned for the code-based encryption schemes, seed expanders can be used to compress keys. The lattice-based cryptosystems that use seed expanders are shown in Table 4 by indicating two key sizes (the key size required when using a seed expander is inside parentheses).

Among the cryptosystems compared in Table 4 that provide a roughly 100-bit quantum security level, it seems that Round5 KEM IoT is the one with the smallest keys and, as it will be later observed in Section V, it provides a fast execution.

#### 4) Supersingular elliptic curve isogeny cryptosystems

These schemes are based on the isogeny protocol for ordinary elliptic curves presented in [124], but enhanced to withstand the quantum attack detailed in [125]. There are different promising post-quantum cryptosystems of this type [126], [127], whose key size is usually in the order of a few thousand bits [128].

Only one isogeny-based public-key encryption scheme passed to the second round of the NIST call: SIKE [129], [130]. SIKE is based on pseudo-random walks in supersingular isogeny graphs. A good reference of SIKE key sizes is SIKEp434, which, for a 128-bit level of classical security, makes use of a 2640-bit public-key and a 2992-bit private key.

#### 5) Hybrid cryptosystems

Hybrid schemes seem to be next step towards post-quantum security, since they merge pre-quantum and post-quantum cryptosystems with the objective of protecting the exchanged data both from quantum attacks and from attacks against the used post-quantum schemes, whose security is currently

**TABLE 4.** Post-quantum lattice-based public-key encryption schemes that passed to the second round of the NIST call.

| Cryptosystem | Subtype | Claimed Quantum Security | Claimed Classical Security | Public Key Size (bits) | Private Key Size (bits) | Key References |
|---|---|---|---|---|---|---|
| CRYSTALS Kyber-512 | Based on solving the LWE problem with Module Lattices | 100 bits | 128 bits | 6,400 | 13,056 (256) | [131], [132], [133], [134] |
| CRYSTALS Kyber-512 90s | Based on solving the LWE problem with Module Lattices | 100 bits | 128 bits | 6,400 | 13,056 (256) | [131], [132], [133], [134] |
| CRYSTALS Kyber-768 | Based on solving the LWE problem with Module Lattices | 164 bits | 192 bits | 9,472 | 19,200 (256) | [131], [132], [133], [134] |
| CRYSTALS Kyber-768 90s | Based on solving the LWE problem with Module Lattices | 164 bits | 192 bits | 9,472 | 19,200 (256) | [131], [132], [133], [134] |
| CRYSTALS Kyber-1024 | Based on solving the LWE problem with Module Lattices | 230 bits | 256 bits | 12,544 | 25,344 (256) | [131], [132], [133], [134] |
| CRYSTALS Kyber-1024 90s | Based on solving the LWE problem with Module Lattices | 230 bits | 256 bits | 12,544 | 25,344 (256) | [131], [132], [133], [134] |
| FrodoKEM-640 AES | Based on solving the LWE problem with generic "algebraically unstructured" lattices | - | 128 bits | 76,928 | 159,104 | [122], [135], [136], [137] |
| FrodoKEM-640 SHAKE | Based on solving the LWE problem with generic "algebraically unstructured" lattices | - | 128 bits | 76,928 | 159,104 | [122], [135], [136], [137] |
| FrodoKEM-976 AES | Based on solving the LWE problem with generic "algebraically unstructured" lattices | - | 192 bits | 125,056 | 250,368 | [122], [135], [136], [137] |
| FrodoKEM-976 SHAKE | Based on solving the LWE problem with generic "algebraically unstructured" lattices | - | 192 bits | 125,056 | 250,368 | [122], [135], [136], [137] |
| FrodoKEM-1344 AES | Based on solving the LWE problem with generic "algebraically unstructured" lattices | - | 256 bits | 172,160 | 344,704 | [122], [135], [136], [137] |
| FrodoKEM-1344 SHAKE | Based on solving the LWE problem with generic "algebraically unstructured" lattices | - | 256 bits | 172,160 | 344,704 | [122], [135], [136], [137] |
| LAC-128 (CCA) | Based on solving Ring-LWE | - | 128 bits | 4,352 | 8,448 | [138] |
| LAC-192 (CCA) | Based on solving Ring-LWE | - | 192 bits | 8,448 | 16,640 | [138] |
| LAC-256 (CCA) | Based on solving Ring-LWE | - | 256 bits | 8,448 | 16,640 | [138] |
| NewHope-512 (CCA) | Based on solving Ring-LWE | 101 bits | 128 bits | 7,424 | 15,104 | [119], [139], [140], [141] |
| NewHope-1024 (CCA) | Based on solving Ring-LWE | 233 bits | 256 bits | 14,592 | 29,440 | [119], [139], [140], [141] |
| NTRUEncrypt (ntruhrss701) | Based on solving LWE/Ring-LWE | 128-192 bits | - | 9,104 | 11,616 | [118], [142], [143] |
| NTRUEncrypt (ntruhps2048677) | Based on solving LWE/Ring-LWE | 128-192 bits | - | 7,448 | 9,880 | [118], [142], [143] |
| NTRUEncrypt (ntruhps4096821) | Based on solving LWE/Ring-LWE | 192-256 bits | - | 9,840 | 12,736 | [118], [142], [143] |
| NTRU Prime (sntrup4591761) | Based on solving Ring-LWE | 139-208 bits | 153-368 bits | 9,744 | 12,800 | [144], [145], [146] |
| NTRU Prime (ntrulpr4591761) | Based on solving Ring-LWE | 140-210 bits | 155-364 bits | 8,376 | 9,904 | [144], [145], [146] |
| Round5 KEM IoT | Based on General Learning with Rounding (GLWR) | 88-101 bits | 96-202 bits | 2,736 | 128 | [147], [148] |
| SABER KEM (LightSABER) | Based on Module Learning With Rounding problem (Mod-LWR) | 114-153 bits | 125-169 bits | 5,376 | 12,544 ( 7,936) | [149], [150] |
| SABER KEM | Based on Module Learning With Rounding problem (Mod-LWR) | 185-226 bits | 203-244 bits | 7,936 | 18,432 (10,752) | [149], [150] |
| SABER KEM (FireSABER) | Based on Module Learning With Rounding problem (Mod-LWR) | 257-308 bits | 283-338 bits | 10,496 | 24,320 (14,080) | [149], [150] |
| Three Bears (BabyBear CCA) | Based on integer module learning with errors (I-MLWE) | 140-180 bits | 154-190 bits | 6,432 | 320 | [151] |
| Three Bears (MamaBear CCA) | Based on integer module learning with errors (I-MLWE) | 213-228 bits | 235-241 bits | 9,552 | 320 | [151] |
| Three Bears (PapaBear CCA) | Based on integer module learning with errors (I-MLWE) | 285-300 bits | 314-317 bits | 12,672 | 320 | [151] |

being evaluated by industry and academia.

This kind of cryptosystems have been tested by Google [152], which merged New Hope [119] with an ECC-based Diffie-Hellman key agreement scheme named X25519. A second version of the hybrid scheme (CECPQ2) is currently being tested: it merges X25519 with instantiations of NTRU (HRSS (Hülsing, Rijneveld, Schanck, Schwabe) and SXY (Saito, Xagawa, Yamakawa)).

Although these schemes look promising, it must be noted that they involve implementing two complex cryptosystems, which require significant computational resources and more energy consumption. Therefore, future developers of hybrid post-quantum cryptosystems for blockchains will have to look for a trade-off between security, computational complexity and resource consumption. In addition, developers will have to address the large payload problem that arises with this kind of cryptosystems when providing Transport Layer Security (TLS) communications (such a problem is due to the required public-key and ciphertext sizes).

## B. POST-QUANTUM SIGNING ALGORITHMS

### 1) Code-based cryptosystems

Different post-quantum code-based signing algorithms have been proposed in the past. Some of the most relevant subtypes of this kind of cryptosystems are based on the schemes from Niederreiter [153] and CFS (Courtois, Finiasz, Sendrier) [154], which are really similar to McEliece's cryptosystem. The signatures of such schemes are short in length and can be verified really fast, but, as it occurs with traditional McEliece's cryptosystems, the use of large key sizes requires significant computational resources and, as a consequence, signature generation may become inefficient.

Other code-based signing algorithms have been proposed in the literature, such as identification protocols related to the application of Fiat-Shamir transformation [155], which in some cases outperform cryptosystems like CFS [156]. Nonetheless, it must be noted that, Fiat-Shamir signatures are not known to be completely secure against quantum attacks [157] (only under certain circumstances [158]), so alternatives like the Unruh transformation should be considered [157].

### 2) Multivariate-based cryptosystems

In this kind of signature schemes the public key is generated through a trapdoor function that acts as private key. This fact usually derives into large public keys, but very small signatures [85].

Some of the most popular multivariate-based schemes rely on Matsumoto-Imai's algorithm, on Isomorphism of Polynomials (IP) [159] or on variants of HFE, which are able to generate signatures with a size comparable to the currently used RSA or ECC-based signatures [160]. Other relevant multivariate-based digital signature schemes have been proposed, like the ones based on pseudo-random multivariate quadratic equations [161] or on Rainbow-like signing schemes (e.g., TTS [162], TRMS [163] or Rainbow [164]). Nonetheless, such cryptosystems need to be further improved in terms of key size, since they usually require several tens of thousands of bytes per key.

Table 5 compares the main characteristics of the digital signature schemes that passed to the second round of the NIST call. In such a Table, for schemes like Rainbow, the values inside parentheses indicate the length of the compressed keys. As it can be observed, among the compared multivariate-based cryptosystems, MQDSS provides really small keys, but the sizes of its signatures are among the largest in the comparison. In contrast, the rest of the compared multivariate-based schemes require several kilobytes for each key, but they produce short signatures (with a length between 239 and 1,632 bits).

### 3) Lattice-based cryptosystems

Among the different lattice-based signature schemes described in the literature, the ones based on Short Integer Solution (SIS) [165] seem to be promising due to their reduced key size. According to some performance analyses, BLISS-B (Bimodal Lattice Signatures B), which relies on the hardness of the SIS problem, provides one of the best performances for lattice-based signing cryptosystems, being on a par with RSA and ECDSA [166]. However, note that the original BLISS [167] was attacked in 2016 under specific conditions through a side-channel attack [168], while its variant BLISS-B is also susceptible to cache attacks that are able to recover the secret signing key after 6,000 signature generations [169].

Besides BLISS, there are in the literature other lattice-based signature schemes that rely on the SIS problem but that were devised specifically for blockchains [170]. Researchers have also developed lattice-based blind signature schemes [171], which were introduced by David Chaum in the early 80s for creating an untraceable payment system [172]. For instance, a lattice-based blind signature scheme is detailed in [173], which was specifically conceived for providing user anonymity and untraceability in distributed blockchain-based applications for IoT.

Finally, it is worth mentioning the lattice-based signature schemes presented in [174], [175]. Specifically, in [174] the authors propose a cryptosystem whose public and private keys are generated through Bonsai Trees [176]. Regarding the work in [175], it presents a lattice-based signature scheme optimized for embedded systems, which, for a 100-bit security level, makes use of a public key of 12,000 bits and a private key of 2,000 bits, and generates signatures of 9,000 bits. This latter scheme, due to its simplicity and efficiency, was selected as signature algorithm for blockchain-related developments like QChain [177], a post-quantum decentralized system for managing public-key encryption.

Table 5 allows for comparing the main characteristics of the lattice-based schemes that passed to the second round of the NIST call. As it can be observed, lattice-based signature schemes require keys whose size is in general smaller than the one needed by multivariate-based schemes, but the generated signatures are slightly larger. Among the com-pared lattice-based cryptosystems, FALCON makes use of the smallest key sizes and signature lengths. Other schemes like qTESLA are fast (as it will be later observed in Section V), but its major drawback is its large key sizes [192].

**TABLE 5.** Post-quantum digital signature schemes that passed to the second round of the NIST call.

| Algorithm | Type | Subtype | Claimed Quantum Security | Public-Key Size | Private Key Size | Signature Size | Key References |
|---|---|---|---|---|---|---|---|
| DILITHIUM 1280x1024 SHAKE (recommended) | Lattice-based | Based on the "Fiat-Shamir with Aborts" technique | 128 bits | 1,472 bytes | - | 2,701 bytes | [178] |
| DILITHIUM 1280x1024 AES (recommended) | Lattice-based | Based on the "Fiat-Shamir with Aborts" technique | 128 bits | 1,472 bytes | - | 2,701 bytes | [178] |
| FALCON-512 | Lattice-based | Based on SIS over NTRU lattices and Fast Fourier sampling | 103 bits | 897 bytes | 1,314.56 (32) bytes | 657.38 bytes | [179], [180] |
| FALCON-1024 | Lattice-based | Based on SIS over NTRU lattices and Fast Fourier sampling | 230 bits | 1,793 bytes | 2,546.62 (32) bytes | 1273.31 bytes | [179], [180] |
| GeMSS 128 | Multivariate-based | Built on HFEv- | 128 bits | 352.19 KB | 13.44 KB | 258 bits | [181], [182] |
| GeMSS 192 | Multivariate-based | Built on HFEv- | 192 bits | 1,237.964 KB | 34.07 KB | 411 bits | [181], [182] |
| GeMSS 256 | Multivariate-based | Built on HFEv- | 256 bits | 3,040.70 KB | 75.89 KB | 576 bits | [181], [182] |
| LUOV Level 1 (Chacha8) | Multivariate-based | Based on Unbalanced Oil and Vinegar (UOV) | 128 bits | 11.5 KB | 32 bytes | 239 bytes | [183], [184] |
| LUOV Level 3 (Chacha8) | Multivariate-based | Based on Unbalanced Oil and Vinegar (UOV) | 192 bits | 35.4 KB | 32 bytes | 337 bytes | [183], [184] |
| LUOV Level 5 (Chacha8) | Multivariate-based | Based on Unbalanced Oil and Vinegar (UOV) | 256 bits | 82.0 KB | 32 bytes | 440 bytes | [183], [184] |
| MQDSS 31-48 | Multivariate-based | Based on the 5-pass SSH (Sakumoto, Shirai, and Hiwatari) identification scheme | 64-128 bits | 46 bytes | 16 bytes | 20,854 bytes | [185], [186] |
| MQDSS 31-64 | Multivariate-based | Based on the 5-pass SSH (Sakumoto, Shirai, and Hiwatari) identification scheme | 96-192 bits | 64 bytes | 24 bytes | 43,728 bytes | [185], [186] |
| PICNIC2 L1-FS | Hash-based | Relies on Zero-Knowledge Proofs | 128 bits | 32 bytes | 16 bytes | 13,802 bytes (max) | [187] |
| PICNIC2 L3-FS | Hash-based | Relies on Zero-Knowledge Proofs | 192 bits | 48 bytes | 24 bytes | 29,750 bytes (max) | [187] |
| PICNIC2 L5-FS | Hash-based | Relies on Zero-Knowledge Proofs | 256 bits | 64 bytes | 32 bytes | 54,732 bytes (max) | [187] |
| qTESLA-p-I | Lattice-based | Based on RLWE | 128 bits | 14,880 bytes | 5,184 bytes | 2,592 bytes | [188] |
| qTESLA-p-III | Lattice-based | Based on RLWE | 192 bits | 38,432 bytes | 12,352 bytes | 5,664 bytes | [188] |
| Rainbow Ia | Multivariate-based | - | 128 bits | 149 bytes (58.1) KB | 93 KB | 512 bits | [189] |
| Rainbow IIIc | Multivariate-based | - | 192 bits | 710.6 (206.7) KB | 511.4 KB | 1,248 bits | [189] |
| Rainbow Vc | Multivariate-based | - | 256 bits | 1,705.5 (491.9) KB | 1,227.1 KB | 1,632 bits | [189] |
| SPHINCS+ -SHAKE256-128f-simple | Hash-based | Stateless signature scheme | 128 bits | 32 bytes | 64 bytes | 16,976 bytes | [190], [191] |
| SPHINCS+ -SHAKE256-192f-simple | Hash-based | Stateless signature scheme | 192 bits | 48 bytes | 96 bytes | 35,664 bytes | [190], [191] |
| SPHINCS+ -SHAKE256-256f-simple | Hash-based | Stateless signature scheme | 256 bits | 64 bytes | 128 bytes | 49,216 bytes | [190], [191] |
| SPHINCS+ -SHA-256-128f-simple | Hash-based | Stateless signature scheme | 128 bits | 32 bytes | 64 bytes | 16,976 bytes | [190], [191] |
| SPHINCS+ -SHA-256-192f-simple | Hash-based | Stateless signature scheme | 192 bits | 48 bytes | 96 bytes | 35,664 bytes | [190], [191] |
| SPHINCS+ -SHA-256-256f-simple | Hash-based | Stateless signature scheme | 256 bits | 64 bytes | 128 bytes | 49,216 bytes | [190], [191] |
| SPHINCS+-Haraka-128f-simple | Hash-based | Stateless signature scheme | 128 bits | 32 bytes | 64 bytes | 16,976 bytes | [190], [191] |
| SPHINCS+-Haraka-192f-simple | Hash-based | Stateless signature scheme | 192 bits | 48 bytes | 96 bytes | 35,664 bytes | [190], [191] |
| SPHINCS+-Haraka-256f-simple | Hash-based | Stateless signature scheme | 256 bits | 64 bytes | 128 bytes | 49,216 bytes | [190], [191] |

### 4) Supersingular elliptic curve isogeny cryptosystems

It is possible to use supersingular elliptic curve isogenies for creating post-quantum digital signature schemes [193], but there are not in the literature many of such schemes and they still suffer from poor performance. For instance, in [194] the authors present different signature schemes based on isogeny problems and on the Unruh transform, which makes use of small key sizes and relatively efficient signing and verification algorithms. Another signature scheme based on the Unruh transform is presented in [195], which, for a 128-bit quantum security level, makes use of a 336-byte public key and a 48-byte private key, but it generates 122,880-byte signatures (even when using compression techniques). Therefore, it is necessary to address key size issues when implementing isogeny-based cryptosystems and Supersingular Isogeny Diffie-Hellman (SIDH), especially in the case of resource-constrained devices, which need to use key compression techniques that often involve computationally intensive steps [196], [197].

### 5) Hash-based signature schemes

The security of these schemes depends on the security of the underlying hash function instead of on the hardness of a mathematical problem. This kind of schemes date back from the late 70s, when Lamport proposed a signature scheme based on a one-way function [198]. Currently, variants of eXtended Merkle Signature Scheme (XMSS) [199] like XMSS-T and SPHINCS [200] are considered promising hash-based signature schemes for the post-quantum era that derive from the Merkle tree scheme described in [201].

However, some researchers consider XMSS and SPHINCS to be impractical for blockchain applications due to their performance [202], so alternatives have been suggested. For example, XMSS has been adapted to blockchain by making use of a single authentication path instead of a tree, while using one-time and limited keys in order to preserve anonymity and minimize user tracking [203]. Other authors [202] proposed substituting XMSS with XNYSS (eXtended Naor-Yung Signature Scheme), a signature scheme that combines a hash-based one-time signature scheme with Naor-Yung chains, which allow for creating chains of related signatures [204].

## V. PERFORMANCE COMPARISON OF POTENTIAL BLOCKCHAIN POST-QUANTUM CRYPTOSYSTEMS
### A. PUBLIC-KEY ENCRYPTION SCHEMES

Tables 6 and 7 compare the post-quantum public-key encryption cryptosystems previously mentioned in Section IV when executed on hardware that can run both a regular blockchain node (i.e., a node that only interacts with the blockchain) or a full blockchain node (i.e., a node that stores and updates periodically a copy of the blockchain and that is able to validate blockchain transactions).

For the sake of fairness, all the evaluation microprocessors indicated in Tables 6 and 7 are based on Intel x64 architecture and had Turbo Boost and Hyper-Threading features disabled. Since the version of the Intel microprocessor varies among the compared cryptosystems, the obtained results should be analyzed considering the differences in microprocessor performance. To carry out such an analysis in a fair way, Table 9 shows the most relevant characteristics of each microprocessor whose performance is referenced in this article. Thus, Table 9 compares the different clock frequencies, the main target platforms (i.e., laptop, server or desktop), the microprocessor typical energy consumption (indicated as Thermal Design Power (TDP)) and the estimated performance (making use of the Passmark CPU benchmarks [205]). In addition, also for the sake of fairness, Tables 6 to 8 compare the obtained performance results on the number of required execution cycles, which means that they have been normalized by taking the specific microprocessor clock frequency into account.

Specifically, Tables 6 and 7 indicate the number of cycles required by each microprocessor for key generation, encapsulation/encryption and decapsulation/decryption. The cycles required by LEDACrypt are not included because in their NIST second-round documentation it is only indicated the total algorithm execution time instead of the number of cycles. For CRYSTALS-Kyber, Table 6 indicates inside the parentheses the estimated number of cycles for the case when key generation is included in the decapsulation process (to avoid having to store expanded private keys).

In order to show in a clear and fast way to the reader which algorithms perform the better on the hardware platforms indicated in Tables 6 and 7 (i.e., without normalizing the performance differences related to the use of different clock frequencies), Figure 2 shows a bar chart of the average execution times of the algorithms listed in such Tables 6 and 7. As it can be observed, the lightest versions of schemes like NTRU Prime, Three Bears and SABER are really fast. However, it is important to note that, while Three Bears and SABER were evaluated in low-power microprocessors for laptops, the results obtained for NTRU Prime were obtained when ran on an Intel Xeon processor, which is a powerful microprocessor for servers.

In contrast, SIKE is the overall slowest scheme among the ones compared, while a cryptosystem like Classic McEliece suffers from a really slow key generation in spite of obtaining reduced decapsulation/decryption and encapsulation/encryption times. Nonetheless, it must be indicated that such slow schemes may be optimized for certain computational architectures and thus provide smallest execution times. In addition, post-quantum schemes, once publicly shared, evolve fast, so new implementations may be released in the future with the objective of reducing their computational complexity and, as a consequence, the required execution time.

**TABLE 6.** Performance comparison of post-quantum encryption algorithms for blockchain nodes (part 1).

| References | Cryptosystem | Claimed Classical Security | Performance Evaluation Hardware | Key Generation (#Cycles) | Encapsulation (#Cycles) | Decapsulation (#Cycles) |
|---|---|---|---|---|---|---|
| [95] | BIKE-1 Level 1 | 128 bits | Intel Core i5-6260U @ 1.80 GHz, 32 GB of RAM | 730,025 | 689,193 | 2,901,203 |
| [95] | BIKE-1 Level 3 | 192 bits | Intel Core i5-6260U @ 1.80 GHz, 32 GB of RAM | 1,709,921 | 1,850,425 | 7,666,855 |
| [95] | BIKE-1 Level 5 | 256 bits | Intel Core i5-6260U @ 1.80 GHz, 32 GB of RAM | 2,986,647 | 3,023,816 | 17,483,906 |
| [95] | BIKE-2 Level 1 | 128 bits | Intel Core i5-6260U @ 1.80 GHz, 32 GB of RAM | 6,383,408 | 281,755 | 2,674,115 |
| [95] | BIKE-2 Level 3 | 192 bits | Intel Core i5-6260U @ 1.80 GHz, 32 GB of RAM | 22,205,901 | 710,970 | 7,114,241 |
| [95] | BIKE-2 Level 5 | 256 bits | Intel Core i5-6260U @ 1.80 GHz, 32 GB of RAM | 58,806,046 | 1,201,161 | 16,385,956 |
| [95] | BIKE-3 Level 1 | 128 bits | Intel Core i5-6260U @ 1.80 GHz, 32 GB of RAM | 433,258 | 575,237 | 3,437,956 |
| [95] | BIKE-3 Level 3 | 192 bits | Intel Core i5-6260U @ 1.80 GHz, 32 GB of RAM | 1,100,372 | 1,460,866 | 7,732,167 |
| [95] | BIKE-3 Level 5 | 256 bits | Intel Core i5-6260U @ 1.80 GHz, 32 GB of RAM | 2,300,332 | 3,257,675 | 18,047,493 |
| [102] | Classic McEliece (mceliece8192128) | 256 bits | Intel Xeon E3-1220 v3 @ 3.10 GHz | ←4,675,000,000 | ←296,000 | ←458,000 |
| [131] | CRYSTALS Kyber-512 | 128 bits | Intel Core i7-4770K @ 3.5 GHz | 118,044 | 161,440 | 190,206 (↑ 279, 150) |
| [131] | CRYSTALS Kyber-512 90s | 128 bits | Intel Core i7-4770K @ 3.5 GHz | 232,368 | 285,336 | 313,452 (↑ 436, 088) |
| [131] | CRYSTALS Kyber-768 | 192 bits | Intel Core i7-4770K @ 3.5 GHz | 217,728 | 272,254 | 315,976 (↑ 469, 008) |
| [131] | CRYSTALS Kyber-768 90s | 192 bits | Intel Core i7-4770K @ 3.5 GHz | 451,018 | 514,088 | 556,972 (↑ 758, 934) |
| [131] | CRYSTALS Kyber-1024 | 256 bits | Intel Core i7-4770K @ 3.5 GHz | 331,418 | 396,928 | 451,096 (↑ 667, 596) |
| [131] | CRYSTALS Kyber-1024 90s | 256 bits | Intel Core i7-4770K @ 3.5 GHz | 735,382 | 810,398 | 860,272 (↑ 1, 148, 394) |
| [135] | FrodoKEM-640 AES | 128 bits | Intel Core i7-6700 @ 3.4 GHz | 1,384,000 | 1,858,000 | 1,749,000 |
| [135] | FrodoKEM-640 SHAKE | 128 bits | Intel Core i7-6700 @ 3.4 GHz | 7,626,000 | 8,362,000 | 8,248,000 |
| [135] | FrodoKEM-976 AES | 192 bits | Intel Core i7-6700 @ 3.4 GHz | 2,820,000 | 3,559,000 | 3,400,000 |
| [135] | FrodoKEM-976 SHAKE | 192 bits | Intel Core i7-6700 @ 3.4 GHz | 16,841,000 | 18,077,000 | 17,925,000 |
| [135] | FrodoKEM-1344 AES | 256 bits | Intel Core i7-6700 @ 3.4 GHz | 4,756,000 | 5,981,000 | 5,748,000 |
| [135] | FrodoKEM-1344 SHAKE | 256 bits | Intel Core i7-6700 @ 3.4 GHz | 30,301,000 | 32,611,000 | 32,387,000 |
| [106] | HQC Level 1 (hqc-128-1) | 128 bits | Intel Core i7- 7820X CPU @ 3.6 GHz, 16 GB of RAM | 110,000 | 190,000 | 310,000 |
| [106] | HQC Level 3 (hqc-192-1) | 192 bits | Intel Core i7- 7820X CPU @ 3.6 GHz, 16 GB of RAM | 190,000 | 330,000 | 510,000 |
| [106] | HQC Level 5 (hqc-256-1) | 256 bits | Intel Core i7- 7820X CPU @ 3.6 GHz, 16 GB of RAM | 270,000 | 470,000 | 690,000 |
| [138] | LAC-128 (CCA) | 128 bits | Intel Core-i7-4770S @ 3.10 GHz, 7.6 GB of RAM | 90,411 | 160,314 | 216,957 |
| [138] | LAC-192 (CCA) | 192 bits | Intel Core-i7-4770S @ 3.10 GHz, 7.6 GB of RAM | 281,324 | 421,439 | 647,030 |
| [138] | LAC-256 (CCA) | 256 bits | Intel Core-i7-4770S @ 3.10 GHz, 7.6 GB of RAM | 267,831 | 526,915 | 874,742 |

**TABLE 7.** Performance comparison of post-quantum encryption algorithms for blockchain nodes (part 2).

| References | Cryptosystem | Claimed Classical Security | Performance Evaluation Hardware | Key Generation (#Cycles) | Encapsulation (#Cycles) | Decapsulation (#Cycles) |
|---|---|---|---|---|---|---|
| [109] | LEDACrypt KEM Level 1 (for two circulant blocks) | 128 bits | Intel i5-6600 @ 3.6 GHz | - | - | - |
| [109] | LEDACrypt KEM Level 3 (for two circulant blocks) | 192 bits | Intel i5-6600 @ 3.6 GHz | - | - | - |
| [109] | LEDACrypt KEM Level 5 (for two circulant blocks) | 256 bits | Intel i5-6600 @ 3.6 GHz | - | - | - |
| [139] | NewHope-512 (CCA) | 128 bits | Intel Core i7-4770K @ 3.5 GHz | 117,128 | 180,648 | 206,244 |
| [139] | NewHope-1024 (CCA) | 256 bits | Intel Core i7-4770K @ 3.5 GHz | 244,944 | 377,092 | 437,056 |
| [142] | NTRUEncrypt (ntruhrss701) | 128/192 bits | Intel Core i7-4770K @ 3.5 GHz | 23,302,424 | 1,256,210 | 3,642,966 |
| [142] | NTRUEncrypt (ntruhps2048677) | 128/192 bits | Intel Core i7-4770K @ 3.5 GHz | 21,833,048 | 1,313,454 | 3,399,726 |
| [142] | NTRUEncrypt (ntruhps4096821) | 192/256 bits | Intel Core i7-4770K @ 3.5 GHz | 31,835,958 | 1,856,936 | 4,920,436 |
| [144] | NTRU Prime (sntrup4591761) | 153-368 bits | Intel Xeon E3-1275 v3 @ 3.5 GHz | 940,852 | 44,788 | 93,676 |
| [144] | NTRU Prime (ntrulpr4591761) | 155-364 bits | Intel Xeon E3-1275 v3 @ 3.5 GHz | 44,948 | 81,144 | 113,708 |
| [112] | NTS-KEM Level 1 | 128 bits | 16-core server with Intel Xeon E5-2667 v2 @ 3.3 GHz, 256 GB of RAM | 39,388,653 | 124,528 | 650,116 |
| [112] | NTS-KEM Level 3 | 192 bits | 16-core server with Intel Xeon E5-2667 v2 @ 3.3 GHz, 256 GB of RAM | 125,672,723 | 396,513 | 1,181,373 |
| [112] | NTS-KEM Level 5 | 256 bits | 16-core server with Intel Xeon E5-2667 v2 @ 3.3 GHz, 256 GB of RAM | 229,357,286 | 532,168 | 2,500,475 |
| [114] | ROLLO-II 128 | 128 bits | Intel Core i7-7820X @ 3.6 GHz, 16 GB of RAM | 9,620,000 | 1,520,000 | 4,960,000 |
| [114] | ROLLO-II 192 | 192 bits | Intel Core i7-7820X @ 3.6 GHz, 16 GB of RAM | 11,040,000 | 2,000,000 | 6,520,000 |
| [114] | ROLLO-II 256 | 256 bits | Intel Core i7-7820X @ 3.6 GHz, 16 GB of RAM | 11,410,000 | 2,390,000 | 7,940,000 |
| [147] | Round5 KEM IoT | 96-202 bits | MacBook Pro 15.1 with Intel Core i7 @ 2.6 GHz | 56,300 | 97,900 | 59,500 |
| [116] | RQC-I | 128 bits | Intel Core i7-7820X @ 3.6 GHz, 16 GB of RAM | 700,000 | 1,300,000 | 6,660,000 |
| [116] | RQC-II | 192 bits | Intel Core i7-7820X @ 3.6 GHz, 16 GB of RAM | 1,120,000 | 2,180,000 | 14,680,000 |
| [116] | RQC-III | 256 bits | Intel Core i7-7820X @ 3.6 GHz, 16 GB of RAM | 1,820,000 | 3,550,000 | 23,200,000 |
| [149] | SABER KEM (LightSABER) | 125-169 bits | Intel Core i5-7200U @ 2.50 GHz | 85,474 | 108,927 | 119,868 |
| [149] | SABER KEM | 203-244 bits | Intel Core i5-7200U @ 2.50 GHz | 163,333 | 196,705 | 215,733 |
| [149] | SABER KEM (FireSABER) | 283-338 bits | Intel Core i5-7200U @ 2.50 GHz | 259,504 | 308,277 | 341,654 |
| [129] | SIKE (SIKEp434) | 128 bits | Intel Core i7-6700 @ 3.4 GHz | 1,047,991,000 | 1,482,681,000 | 1,790,304,000 |
| [151] | Three Bears (BabyBear CCA) | 154-190 bits | Intel Core i3-6100U @ 2.3 GHz | 41,000 | 60,000 | 101,000 |
| [151] | Three Bears (MamaBear CCA) | 235-241 bits | Intel Core i3-6100U @ 2.3 GHz | 79,000 | 96,000 | 156,000 |
| [151] | Three Bears (PapaBear CCA) | 314-317 bits | Intel Core i3-6100U @ 2.3 GHz | 118,000 | 145,000 | 211,000 |

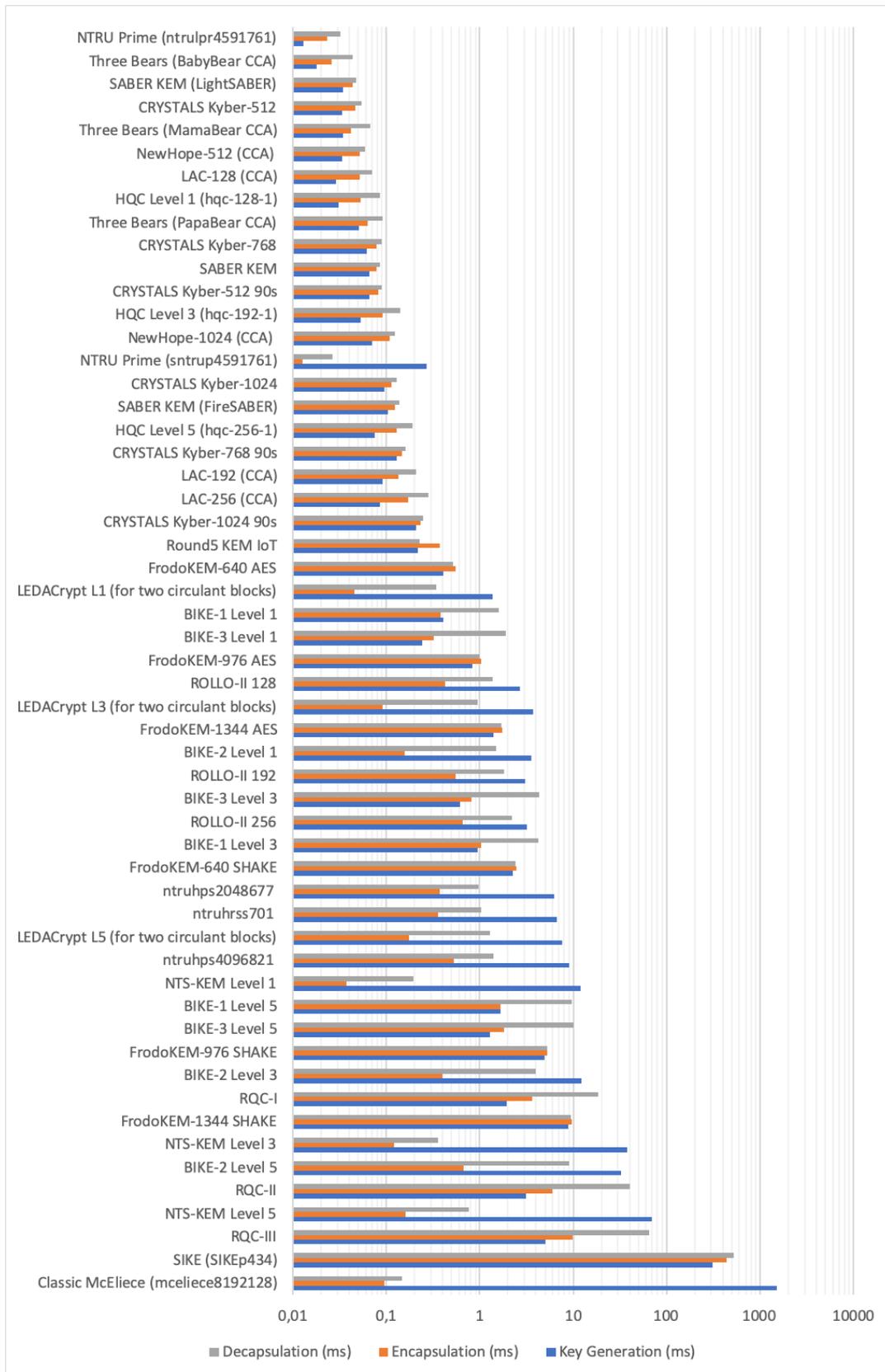

**FIGURE 2.** Comparison of the average execution times (in milliseconds) of NIST call second round public-key encryption schemes.

### B. DIGITAL SIGNATURE SCHEME PERFORMANCE

Table 8 compares the performance of post-quantum digital signature algorithms that passed to the second round of the NIST call. The following considerations should be taken into account regarding the information shown in the Table:

- In the case of FALCON, the authors measured its performance in terms of spent time instead of cycles. This is related to the fact that the processor used by the researchers implemented dynamic frequency scaling based on load and temperature, which derived into measurements that vary up to 15% [180].
- For Rainbow, the values inside the parentheses indicate the performance of the key-compressed version, which, as it can be observed, requires much more computational effort than the regular version due to the involved decompression process.
- Most cryptosystems have been evaluated after optimizing them for AVX2, a 256-bit instruction set provided by Intel. The only exception is SPHINCS+ performance for the HARAKA version, whose optimized version was implemented to take advantage of the AES-NI instruction set.

Figure 3 shows through a bar chart the average execution times for the post-quantum cryptosystems listed in Table 8. Like in the case of the results obtained for the post-quantum encryption schemes, it is worth noting that the compared execution times were obtained in similar but not identical hardware platforms, so performance differences should be considered just as estimations. In addition, the following aspects should be taken into account regarding Figure 3:

- The obtained results are sorted by the sum of the three compared times, which is an estimation of the overall speed of each algorithm.
- FALCON is not included since there are no data for the three compared parameters.
- Besides post-quantum cryptosystems, the time required by two comparable pre-quantum schemes have been included as a reference: ECDSA (P-256) and RSA-3072. The execution times shown in the Figure for such implementations were obtained from [166], where the author used the libstrongswan library, which acted as an openssl wrapper for RSA and ECDSA, and whose measurements were performed on a laptop with an Intel Core i7-3610QM CPU at 2.30 GHz.
- The obtained results show that, as it was expected, the AVX2/AES-NI optimizations are clearly faster than the reference versions.
- The fastest schemes are DILITHIUM and the lightest versions of LUOV, qTESLA, MQDSS and Rainbow. Overall, the AVX2 optimization of DILITHIUM seems to be, in terms of execution time, the most promising post-quantum digital signature scheme, since it obtains very similar results to ECDSA-256. Unfortunately, DILITHIUM key sizes are larger than the ones used by ECDSA-256, so researchers should focus on developing new approaches to reduce them.
- The slowest cryptosystems are the most secure versions of PICNIC2, GeMSS, Rainbow and SPHINCS. In the case of PICNIC2, its lack of speed is due to slow verification and signing processes. Regarding GeMSS, Rainbow and SPHINCS, their execution speed is impacted significantly by the amount of time devoted to key generation.

## VI. POST-QUANTUM BLOCKCHAIN PROPOSALS

Different authors have already proposed post-quantum blockchains or modifications of current blockchains to tackle the quantum threat [206], [207], [208]. For instance, in [209] it is proposed a framework aimed at sharing sensitive industrial data in public distributed networks. Such a framework is able to work with Inter-Planetary File System (IPFS) and Ethereum, and implements Diffie-Hellman Key Exchange on SIDH. Ethereum is also modified in [210], but with the multivariate-based cryptosystem Rainbow, whose performance is compared in the cited article with the current Ethereum version (based on ECDSA).

In the case of [211], the authors propose to improve Bitcoin (which uses the Koblitz curve secp256k1 and SHA-256 during the ECDSA signature process) with TESLA# [212], which makes use of BLAKE2 [213] and SHA-3 [214]. It is also worth mentioning the work in [10], where it is presented a blockchain-based transparent e-voting protocol that makes use of Niederreiter's code-based cryptosystem to proof the system against quantum attacks.

Other authors have suggested the implementation of quantum-safe blockchains [39], [215]. For example, in [215] the researchers present a quantum-safe transaction authentication scheme based on lattice-based cryptography and provide a standard transaction model to prevent quantum attacks. Similarly, in [39] a lattice-based signature scheme is proposed for developing a post-quantum blockchain that can be used to implement a cryptocurrency.

Commercial blockchains have also analyzed and addressed the impact of quantum computers. DLTs like IOTA's Tangle [40] claim to be more resistant than Bitcoin to quantum attacks that affect processes like nonce search [216]. In addition, IOTA has the advantage of being based on one-time hash-based signatures (Winternitz signatures) instead of on ECC. Furthermore, IOTA is expected to make use of ternary hardware (instead of traditional binary hardware) that will implement a new hash function called CURL-P, which is currently being audited. Finally, it is worth mentioning that there are other blockchains that have been devised to replace Bitcoin in the post-quantum era, like Quantum-Resistant Ledger [217], which replaces secp256k1 with XMSS.

## VII. MAIN CHALLENGES AND FUTURE RESEARCH TOPICS IN POST-QUANTUM BLOCKCHAIN

### A. QUANTUM COMPUTING FAST EVOLUTION

Quantum computing is currently a hot topic that has attracted a lot of attention from academia and industry. As a con-

**TABLE 8.** Performance comparison of post-quantum digital signature algorithms that passed to the second round of the NIST call.

| | | Performance (Reference Implementation) | | | Performance (Optimized AVX2 Implementation) | | |
|---|---|---|---|---|---|---|---|
| Algorithm | Evaluation Platform | Key Generation (#Cycles) | Signing (#Cycles) | Verification (#Cycles) | Key Generation (#Cycles) | Signing (#Cycles) | Verification (#Cycles) |
| DILITHIUM 1280x1024 SHAKE (recommended) | Intel Core-i7 6600U (Skylake) CPU @ 2.6 GHz | 371,083 | 1,562,215 | 375,708 | 156,777 | 437,638 | 155,784 |
| DILITHIUM 1280x1024 AES (recommended) | Intel Core-i7 6600U (Skylake) CPU @ 2.6 GHz | - | - | - | 99,907 | 350,465 | 109,782 |
| FALCON-512 | Intel Core i7-6567U @ 3.3 GHz | 7.26 ms | - | - | - | - | - |
| FALCON-1024 | Intel Core i7-6567U @ 3.3 GHz | 21.63 ms | - | - | - | - | - |
| GeMSS 128 | Intel Core i7-6600U CPU @ 2.60 GHz (Skylake) | - | - | - | 36,800,000 | 529,000,000 | 84,600,000 |
| GeMSS 192 | Intel Core i7-6600U CPU @ 2.60 GHz (Skylake) | - | - | - | 167,000,000 | 1,720,000,000 | 233,000,000 |
| GeMSS 256 | Intel Core i7-6600U CPU @ 2.60 GHz (Skylake) | - | - | - | 508,000,000 | 2,830,000,000 | 550,000,000 |
| LUOV Level 1 (Chacha8) | Intel Core i5-8250U CPU @ 1.60 GHz | - | - | - | 1,100,000 | 224,000 | 49,000 |
| LUOV Level 3 (Chacha8) | Intel Core i5-8250U CPU @ 1.60 GHz | - | - | - | 4,600,000 | 643,000 | 152,000 |
| LUOV Level 5 (Chacha8) | Intel Core i5-8250U CPU @ 1.60 GHz | - | - | - | 9,700,000 | 1,100,000 | 331,000 |
| MQDSS 31-48 | Intel Core i7-4770K CPU @ 3.5 GHz | 1,192,984 | 26,630,590 | 19,840,136 | 1,074,644 | 3,816,106 | 2,551,270 |
| MQDSS 31-64 | Intel Core i7-4770K CPU @ 3.5 GHz | 2,767,384 | 85,268,712 | 62,306,098 | 2,491,050 | 9,047,148 | 6,132,948 |
| PICNIC2 L1-FS | Intel Core i7-4790 CPU @ 3.60 GHz | 149,749 | 3,066,663,719 | 1,857,340,295 | 21,026 | 229,947,918 | 100,546,772 |
| PICNIC2 L3-FS | Intel Core i7-4790 CPU @ 3.60 GHz | 362,481 | 10,190,171,124 | 5,537,696,230 | 20,160 | 657,944,759 | 223,785,326 |
| PICNIC2 L5-FS | Intel Core i7-4790 CPU @ 3.60 GHz | 691,790 | 25,488,037,138 | 12,943,455,830 | 35,716 | 1,346,724,260 | 387,637,876 |
| qTESLA-p-I | Intel Core i7-6700 (Skylake) @ 3.4 GHz | 2,316,200 | 2,324,900 | 671,400 | - | - | - |
| qTESLA-p-III | Intel Core i7-6700 (Skylake) @ 3.4 GHz | 13,726,600 | 6,284,600 | 1,830,400 | - | - | - |
| Rainbow Ia | Reference: Intel Xeon CPU E3-1225 v5 @ 3.30 GHz (Skylake), AVX2: Intel Xeon CPU E3-1275 v5 @ 3.60 GHz (Skylake) | 35,000,000 (40,200,000) | 402,000 (20,200,000) | 155,000 (3,440,000) | 8,290,000 (9,280,000) | 67,700 (6,410,000) | 21,700 (3,370,000) |
| Rainbow IIIc | Reference: Intel Xeon CPU E3-1225 v5 @ 3.30 GHz (Skylake), AVX2: Intel Xeon CPU E3-1275 v5 @ 3.60 GHz (Skylake) | 340,000,000 (402,000,000) | 1,700,000 (217,000,000) | 1,640,000 (19,400,000) | 94,800,000 (110,000,000) | 588,000 (61,800,000) | 114,000 (17,800,000) |
| Rainbow Vc | Reference: Intel Xeon CPU E3-1225 v5 @ 3.30 GHz (Skylake), AVX2: Intel Xeon CPU E3-1275 v5 @ 3.60 GHz (Skylake) | 757,000,000 (879,000,000) | 3,640,000 (469,000,000) | 2,390,000 (45,400,000) | 126,000,000 (137,000,000) | 755,000 (87,200,000) | 197,000 (4,300,0000) |
| SPHINCS+ -SHAKE256-128f-simple | Intel Core i7-4770K CPU @ 3.5 GHz | 10,829,190 | 350,847,594 | 13,922,112 | 3,909,682 | 133,452,230 | 9,468,278 |
| SPHINCS+ -SHAKE256-192f-simple | Intel Core i7-4770K CPU @ 3.5 GHz | 15,192,014 | 645,965,282 | 21,943,196 | 6,303,298 | 171,354,532 | 14,758,202 |
| SPHINCS+ -SHAKE256-256f-simple | Intel Core i7-4770K CPU @ 3.5 GHz | 74,279,484 | 902,307,648 | 21,261,734 | 16,898,344 | 416,398,690 | 15,383,888 |
| SPHINCS+ -SHA-256-128f-simple | Intel Core i7-4770K CPU @ 3.5 GHz | 15,426,726 | 693,497,446 | 13,449,776 | 3,257,486 | 116,197,711 | 6,094,962 |
| SPHINCS+ -SHA-256-192f-simple | Intel Core i7-4770K CPU @ 3.5 GHz | 21,274,744 | 464,737,100 | 20,803,660 | 2,280,172 | 140,223,132 | 9,723,976 |
| SPHINCS+ -SHA-256-256f-simple | Intel Core i7-4770K CPU @ 3.5 GHz | 71,620,636 | 1,092,969,048 | 22,716,202 | 5,594,338 | 145,433,610 | 9,384,544 |
| SPHINCS+-Haraka-128f-simple | Intel Core i7-4770K CPU @ 3.5 GHz | 21,556,006 | 378,800,946 | 13,712,542 | 654,294 | 25,178,368 | 1,333,172 |
| SPHINCS+-Haraka-192f-simple | Intel Core i7-4770K CPU @ 3.5 GHz | 19,985,722 | 484,198,114 | 44,676,162 | 2,317,102 | 58,491,132 | 3,714,942 |
| SPHINCS+-Haraka-256f-simple | Intel Core i7-4770K CPU @ 3.5 GHz | 82,842,862 | 1,046,811,244 | 20,879,946 | 2,510,894 | 65,870,866 | 1,949,510 |

**TABLE 9.** Specifications of Intel microprocessors used for evaluating potential blockchain post-quantum algorithms.

| Microprocessor | Clock Frequency | Market Segment | Microarchitecture | Typical TDP | Release Data | Passmark Average Mark | Passmark Single Thread Rating |
|---|---|---|---|---|---|---|---|
| Intel Core i3-6100U | 2.3 GHz | Laptop | Skylake | 15 W | Q4 2015 | 3,603 | 1,302 |
| Intel Core i5-6260U | 1.8 GHz | Laptop | Skylake | 15 W | Q4 2015 | 4,362 | 1,593 |
| Intel Core i5-7200U | 2.5 GHz | Laptop | Kaby Lake | 15 W | Q4 2016 | 4,602 | 1,722 |
| Intel Core i7-6600U | 2.6 GHz | Laptop | Skylake | 15 W | Q3 2015 | 4,802 | 1,805 |
| Intel Core i7-6567U | 3.3 GHz | Laptop | Skylake | 28 W | Q3 2015 | 5,582 | 1,984 |
| Intel Xeon E3-1220 v3 | 3.1 GHz | Server | Haswell | 80 W | Q1 2011 | 7,163 | 1,945 |
| Intel Core i5-8250U | 1.6 GHz | Laptop | Kaby Lake | 15 W | Q3 2017 | 7,645 | 1,926 |
| Intel Core i5-6600 | 3.3 GHz | Desktop | Skylake | 65 W | Q2 2015 | 7,778 | 2,095 |
| Intel Xeon E3-1225 v5 | 3.3-3.7 GHz | Server | Skylake | 80 W | Q4 2015 | 7,829 | 1,969 |
| Intel Core-i7-4770S | 3.1 GHz | Desktop | Haswell | 65 W | Q2 2013 | 9,348 | 2,174 |
| Intel Xeon E3-1275 v3 | 3.5 GHz | Server | Haswell | 95 W | Q2 2013 | 9,915 | 2,214 |
| Intel Core i7-4790 | 3.6 GHz | Desktop | Haswell | 84 W | Q2 2014 | 9,989 | 2,283 |
| Intel Core i7-6700 | 3.4 GHz | Desktop | Skylake | 65 W | Q2 2015 | 10,003 | 2,154 |
| Intel Core i7-4770K | 3.5 GHz | Desktop | Haswell | 84 W | Q2 2013 | 10,075 | 2,250 |
| Intel Core i7-7820X | 3.6 GHz | Desktop | Skylake | 140 W | Q2 2017 | 18,489 | 2,401 |
| Intel Xeon E5-2667 v2 | 3.3 GHz | Server | Sandy Bridge | 130 W | Q1 2014 | 22,568 | 2,023 |

sequence, it is possible that new attacks will be developed against the post-quantum cryptosystems mentioned in this article, so researchers will have to pay attention to the quantum computing scene and its advances.

### B. TRANSITION FROM PRE-QUANTUM TO POST-QUANTUM BLOCKCHAIN

The transition from pre-quantum to post-quantum blockchains requires to think carefully the involved steps. For such a purpose, different researchers have devised methods. For instance, in [218] the authors propose a scheme to extend the validity of past blockchain blocks when the security of a hash function or of the digital signatures is compromised. However, the transition scheme may actually imply a hard-fork of the blockchain, but, to avoid it, a soft-fork mechanism may be implemented [219]. Another mechanism is proposed in [220], where it is presented a simple commit–delay–reveal protocol that enables blockchain users to move in a secure way funds from pre-quantum Bitcoin to a version that implements a post-quantum digital signature scheme.

### C. LARGE KEY AND SIGNATURE SIZES

In general, post-quantum cryptosystems require to use keys whose size is much larger than current public-key cryptosystems (usually between 128 and 4,096 bits).

In the case of digital signature cryptosystems, there are schemes like the ones based on supersingular isogenies that seem promising in terms of key size, but they produce large signatures and its performance is poor in comparison to other cryptosystems. For instance, as it was previously mentioned in Section IV-B4, the scheme detailed in [195], for a 128-bit quantum security level, makes use of 2,688-bit public keys and 384-bit private keys, but it produces signatures of 120 KB, which is a problem for structures like blockchains that have to store massive amounts of such signatures. Similarly, hash-based schemes have a relatively small public/private key size, but their signatures often exceed 40 KB [60]. In contrast, some multivariate-based are able to provide short signatures, but the keys used for generating and verifying such signatures can occupy several kilobytes. Regarding lattice-based schemes, there are versions of DILITHIUM that are really fast, but whose key size is roughly 1,500 bytes and their signature length occupies 2,701 bytes.

With respect to post-quantum public-key encryption cryptosystems, certain optimized versions of schemes like Round5 seem promising, since their performance is good enough for most current blockchain node hardware, while keeping key size low (2,736 bits for the public key and only 128 bits for the private key). Nonetheless, more research is still needed in post-quantum schemes in order to provide a good trade-off between key sizes and security for blockchains.

### D. SLOW KEY GENERATION

In order to increase security, some post-quantum schemes limit the number of messages signed with the same key. As a consequence, it is necessary to generate new keys continuously, which involves dedicating computational resources and slowing down certain blockchain processes. Therefore, blockchain developers will have to determine how to adjust such key generation mechanisms to optimize the blockchain efficiency.

### E. COMPUTATIONAL AND ENERGY EFFICIENCY

As it can be concluded from the comparisons shown in Sections IV and V, some post-quantum schemes require a significant execution time, storage and computational resources. Such needs often derive into increased energy consumption, so future developers will have to look for novel approaches to optimize cryptosystems in order to maximize their com-putational and energy efficiency, and, as a consequence, the efficiency of the overall blockchain.

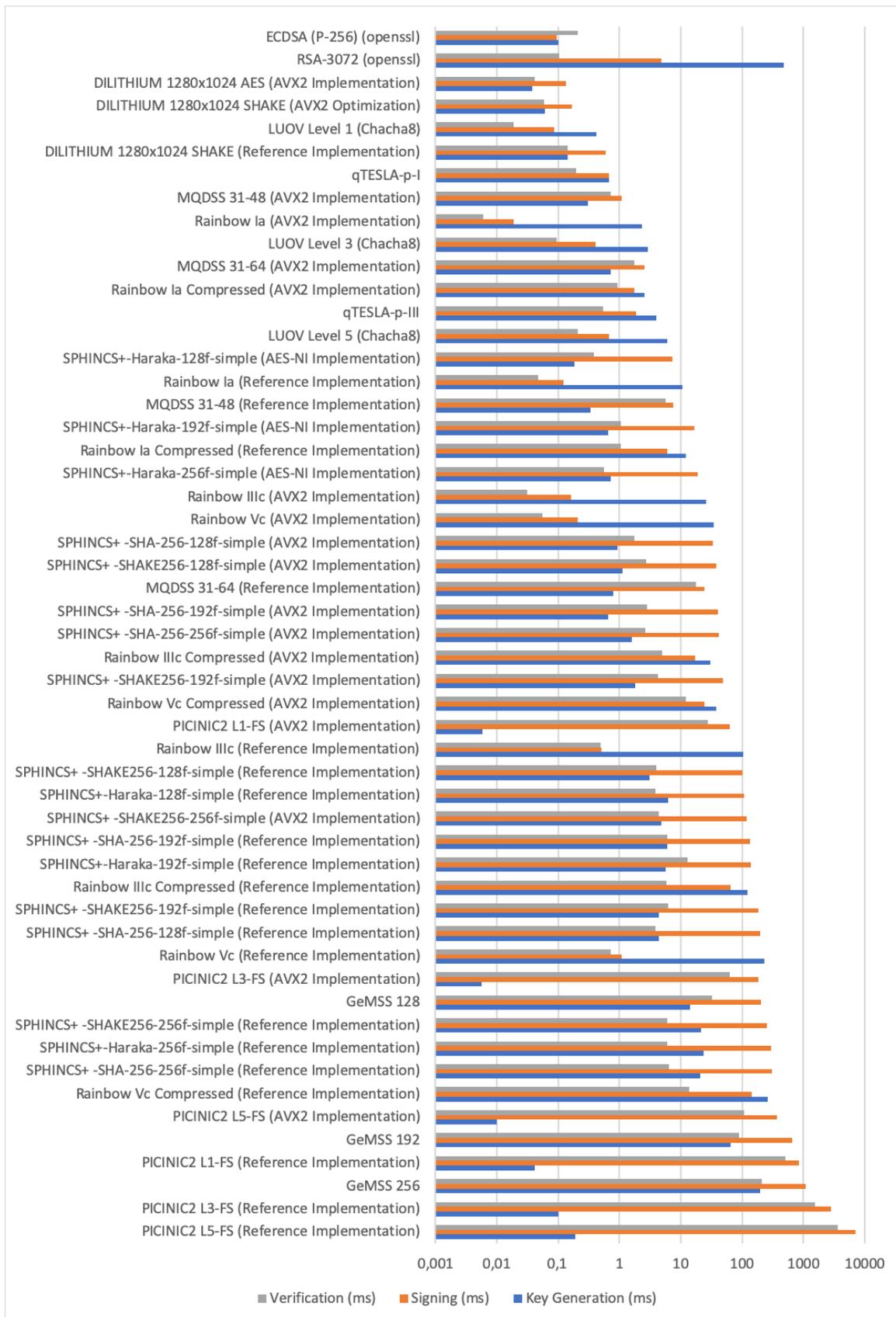

**FIGURE 3.** Comparison of the average execution times (in milliseconds) of NIST call second round digital signature schemes.

*F. STANDARDIZATION*

As it was mentioned in Section III-C, multiple initiatives are currently analyzing post-quantum cryptosystems in order to standardize them. Since this is an ongoing effort, the researchers that look for guaranteeing blockchain compatibility will have to monitor the post-quantum scene and avoid the risk of using non-standard, discarded or broken schemes.

*G. BLOCKCHAIN HARDWARE UNSUITABILITY*

Some computationally intensive post-quantum cryptosystems may not be suitable for certain hardware that is currently used for implementing blockchain nodes. Therefore, post-quantum schemes should provide a trade-off between security and computational complexity so that not to restrict the potential hardware that may interact with the blockchain.

*H. LARGE CIPHERTEXT OVERHEADS*

Certain cryptosystems generate large overheads that may impact the performance of a blockchain. To tackle this issue, future post-quantum developers will have to minimize ciphertext overhead and consider potential compression techniques.

*I. QUANTUM BLOCKCHAIN*

Besides the use of cryptosystems to transition from pre-quantum to post-quantum blockchain, several researchers proposed quantum-computing based blockchains [221], [222], [223]. For instance, in [224] and [225], the authors propose to migrate Bitcoin to quantum computers, while others described how to accelerate mining by modifying Grover's algorithm [226]. Moreover, some authors have already suggested using quantum cryptography to implement smart contracts [227]. Furthermore, more research is necessary on key establishment physics-based methods that are collectively known as Quantum-Key Distribution (QKD) [61].

## VIII. KEY FINDINGS

After the thorough literature analysis carried out in this article, the following conclusions can be drawn:

- After revising the literature, it was found no previous paper that provides a broad view on the importance and application of post-quantum blockchain as it is provided in this article. Although there are other reviews that addressed the impact of quantum computing on blockchain, they were essentially focused on giving generic recommendations for quantum-proofing blockchain [60] or on specific fields [228]. Moreover, it was found no other review that included the following main contributions together:
    - A detailed analysis on the impact of quantum attacks on blockchain public-key cryptosystems and hash functions.
    - A review on the most relevant post-quantum blockchain projects and standardization initiatives.
    - A detailed analysis on the characteristics of the main types of post-quantum encryption and digital signature schemes that can potentially be applied to blockchain.
    - Thorough comparisons on the performance of the most promising post-quantum blockchain cryptosystems.
    - A summary on the main post-quantum blockchain challenges and future trends that will provide a guide for future researchers and developers.
- Although there have been large projects on post-quantum security, it was not found any large academic initiative on the application of such a kind of security to blockchain.
- Nowadays, there are no post-quantum blockchain algorithms that provide, at the same time, small key size, short signature/hash sizes, fast execution, low computational complexity and low energy consumption. Such factors are especially critical for resource-constrained embedded devices like the ones used in the Internet of Things [228].
- Most of the post-quantum cryptosystems whose performance was compared in this article are currently being analyzed by the cryptographic community with the objective of selecting the most appropriate to be standardized through the NIST public call. Therefore, future developers should monitor the news and reports from NIST before selecting a specific post-quantum algorithm.
- It is not straightforward to choose a blockchain post-quantum cryptosystem. Future developers will have to take such a decision based on their blockchain node hardware, on the available resources (i.e., memory, speed), on the required blockchain node performance and on the necessary security level. For such a purpose, the tables provided throughout this article can be a very useful guide to estimate which may be the most promising candidates. Nonetheless, it has to be emphasized that the results provided in this article are related to specific hardware platforms, so performance will vary significantly when implemented and optimized for other hardware.
- Regarding the specific implementations compared in this article, the following general assessments can be stated on their application to blockchain:
    - Coded-based cryptosystems make use of large keys whose management and operation require a relevant amount of computational resources. More research is necessary on key compression techniques and on the use of certain types of codes and coding techniques.
    - Lattice-based cryptosystems also need to be enhanced in terms of key size, but it can be stated

that they are currently some of the most promising candidates for implementing schemes for post-quantum blockchains. In fact, the comparisons performed in this article have shown that lattice-based algorithms Three Bears and SABER are really fast, even when executed on low-power microprocessors for laptops. In addition, a scheme like Round5 KEM IoT seems appropriate for being executed in most current blockchain node hardware and in many applications that do not require very high security. Furthermore, lattice-based digital signature cryptosystems have already been suggested and tested in different practical blockchain implementations [170], [173], [177] and, according to the comparisons shown in this article, certain optimized versions of DILITHIUM and qTESLA are among the fastest ones.
- Multivariate-based public-key cryptosystems still need to be improved to increase decryption speed and to decrease key size. However, it should be noted that some multivariate-based signature algorithms optimized for the AVX2 instruction set (i.e., LUOV, MQDSS and Rainbow) are clearly faster than most of the compared digital signature cryptosystems.
- Hybrid schemes like the ones tested by Google (CECPQ1 and CECPQ2) seem to be the next step prior to the actual implementation of pure post-quantum blockchains, but they require to make use of hardware able to handle at the same time two advanced security mechanisms and large payloads.
- Super-singular elliptic-curve isogenie cryptosystems based on the Unruh transform seem promising, but still need to be optimized to decrease their signature size.
- Hash-based digital signature cryptosystems have in general poor performance, but some researchers have suggested new faster algorithms that seem to be practical for blockchain [203], [202].

- It is necessary to study further how to enhance blockchain security by adding certain features that have been barely used in non-academic blockchain developments and validate their security in the post-quantum era. Some of such features are:
  - Aggregate signatures. They allow for generating a unique signature from several of them. This concept is attractive for blockchain, since it enables faster verification and reduces storage and bandwidth [229].
  - Ring signatures. They allow for specifying a set of possible signers without revealing who of them actually produced a signature [230]. Some researchers have already suggested quantum-resistant lattice-based schemes to secure ring signatures [231], [232], [233] and applied them in blockchain developments [234].
  - Identity-Base Encryption (IBE). It enables a sender and a receiver to communicate without exchanging public or private keys. For such a purpose, a trusted third-party is used as a middle-man between the sender and the receiver to generate private keys, which are sent to the receiver upon request. The scheme has been also generalized as Identity-based Broadcast Encryption (IBBE), which is able to manage multiple receivers instead of only one. IBE and IBBE are interesting for closed groups of users like private blockchains [235] and there are already implementations [236] (even for embedded systems [237]), but their need for a trusted third-party seems to be in conflict with the concept of public blockchain, whose existence is precisely justified by the lack of trust.
  - Secret sharing. It consists in dividing a piece of sensitive information into multiple parts that are distributed among diverse participants and which can be reconstructed by using a minimum number of parts [238]. For instance, in [8] it is introduced a private-key distribution method to help recover lost private keys that is based in secret sharing and in network protocols that guarantee the security of secret share transmission. Another example can be found in [239], where the authors use secret sharing to distribute transaction data securely among peers in a blockchain.
  - Homomorphic encryption. It enables third-party services to process a transaction without revealing unencrypted data to them [240], [241]. This kind of encryption has been already proposed to enhance the Bitcoin protocol [242], [243] and for blockchain-based IoT systems [244].
  - Zero-Knowledge Proofs. This kind of proofs validate a statement without revealing any secret related to it [245]. There is a specific type of these proofs called Zero-Knowledge Succinct Non-Interactive Argument of Knowledge (zk-SNARK) that is aimed at reducing the complexity and the size of the proof [246]. However, it is necessary to design zk-SNARKs to make use of post-quantum cryptosystems or to take advantage of new post-quantum schemes like zk-STARKs [247]. In addition, it is possible to make use of SNAGS (Succinct Non-Interactive Arguments), whose quantum-resistivity is still being studied by the research community [248].
  - Secure Multi-Party Computation (SMPC). SMPC allows the parties involved in a blockchain to act together, but in a way that a single party does not have access to all the information, thus preventing secret data leaks. An example of the use of SMPC on a blockchain is Enigma [249], which first stores hashes on a blockchain and then the related data on *i* Although the analyses carried out in this article are focused on blockchain, since other DLTs work in a similar way, it is quite straightforward to apply to them the provided recommendations and extracted conclusions. Thus, such recommendations and conclusions could be extrapolated to DLTs based on Directed Acyclic Graphs (DAGs) (e.g., IOTA [40], Byteball [250]) or on Hashgraphs (e.g., Swirlds [251]). However, researchers still need to evaluate thoroughly DLT implementations that have already claimed to be better prepared for the post-quantum era than certain blockchains (e.g., IOTA, Quantum-Resistant Ledger [217]).

## IX. CONCLUSION

The recent progress on quantum computing has sparked interest in researchers and developers that work with DLTs like blockchain, where public-key cryptography and hash functions are essential. This article analyzed the impact of quantum-computing attacks (based on Grover's and Shor's algorithms) on blockchain and studied how to apply post-quantum cryptosystems to mitigate such attacks. For such a purpose, the most relevant post-quantum schemes were reviewed and their application to blockchain was analyzed, as well as their main challenges. In addition, extensive comparisons were provided on the characteristics and performance of the most promising post-quantum public-key encryption and digital-signature schemes. Thus, this article gives a broad view and insights on the quantum threat on blockchain, and provides useful guidelines for the researchers and developers of the next-generation of quantum-resistant blockchains.

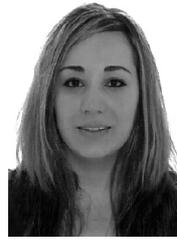

PAULA FRAGA-LAMAS (M'17) received the M.Sc. degree in Computer Engineering from the University of A Coruña (UDC) in 2009, and the M.Sc. and Ph.D. degrees in the joint program Mobile Network Information and Communication Technologies from five Spanish universities: University of the Basque Country, University of Cantabria, University of Zaragoza, University of Oviedo and University of A Coruña, in 2011 and 2017, respectively. Since 2009, she has been with the Group of Electronic Technology and Communications (GTEC), Department of Computer Engineering (UDC). She holds an MBA and postgraduate studies in business innovation management (Jean Monnet Chair in European Industrial Economics, UDC), Corporate Social Responsibility (CSR) and social innovation (Inditex-UDC Chair of Sustainability). She has over 50 contributions in indexed international journals, conferences, and book chapters, and four patents. Her current research interests include mission-critical scenarios, Industry 4.0, Internet of Things (IoT), Cyber-Physical Systems (CPS), Augmented Reality (AR), blockchain and Distributed Ledger Technology (DLT), and cybersecurity. She has also been participating in over 30 research projects funded by the regional and national government as well as the R&D contracts with private companies.

...

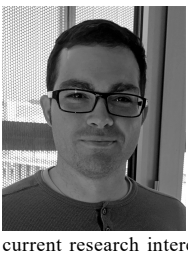

TIAGO M. FERNÁNDEZ-CARAMÉS (S'08-M'12-SM'15) works since 2016 as an Assistant Professor in the area of Electronic Technology at the University of A Coruña (UDC) (Spain), where he obtained his MSc degree and PhD degrees in Computer Science in 2005 and 2011. Since 2005 he has worked in the Department of Computer Engineering at UDC: from 2005 to 2009 through different predoctoral scholarships and between 2007 and 2016 as Interim Professor. His current research interests include IoT/IIoT systems, RFID, wireless sensor networks, augmented reality, embedded systems and blockchain, as well as the different technologies involved in the Industry 4.0 paradigm. In such fields, he has contributed to 40 papers for conferences, to 35 articles for JCR-indexed journals and to two book chapters. Due to his expertise in the previously mentioned fields, he has acted as peer reviewer and guest editor for different top-rank journals, and as project reviewer for national research bodies from Austria (FWF), Croatia (CSF) and Argentina (ANCyT).